\documentclass[aps,prc,twocolumn,groupedaddress,showpacs,superscriptaddress,nofootinbib]{revtex4-1}
\usepackage{graphicx,amsmath}
\usepackage{amssymb}
\usepackage{draftwatermark}

\begin{document}
\SetWatermarkScale{0}

\title{Calculation of resonance energies from Q-values 
}



\author{Christian Iliadis}
\affiliation{Department of Physics \& Astronomy, University of North Carolina at Chapel Hill, NC 27599-3255, USA}
\affiliation{Triangle Universities Nuclear Laboratory (TUNL), Durham, North Carolina 27708, USA}

\date{\today}

\begin{abstract}
Resonance energies are frequently derived from precisely measured excitation energies and reaction Q-values. The latter quantities are usually calculated from atomic instead of nuclear mass differences. This procedure disregards the energy shift caused by the difference in the total electron binding energies before and after the interaction. Assuming that the interacting nuclei in a stellar plasma are fully ionized, this energy shift can have a significant effect, considering that the resonance energy enters exponentially into the expression for the narrow-resonance thermonuclear reaction rates. As an example, the rate of the $^{36}$Ar(p,$\gamma$)$^{37}$K reaction is discussed, which, at temperatures below $1$~GK, depends only on the contributions of a single resonance and direct capture. In this case, disregarding the energy shift caused by the total electron binding energy difference erroneously enhances the rate by $\approx$40\% near temperatures of $70$~MK.
\end{abstract}

\pacs{}

\maketitle


\section{Introduction} \label{sec:intro}
Knowledge of precise center-of-mass resonance energies is essential for estimating thermonuclear reaction rates \cite{iliadis15}. Resonance energies can be deduced, for example, from measured thick- or thin-target excitation functions. This procedure requires accurate knowledge of the beam energy, beam resolution, and target surface composition. For this reason, most resonance energies, especially for narrow resonances, are calculated from well-known excitation energies, $E_x$, in the compound nucleus and the reaction Q-value, according to $E_{cm}$ $=$ $E_x$ $-$ $Q$. Excitation energies of unbound states are usually well-known, with an uncertainty of a fraction of a kiloelectronvolt. Since many Q-values can be computed precisely from evaluated masses, the deduced resonance energies have small uncertainties.  

The purpose of this report is to draw attention to the widespread practice of adopting {\it atomic} masses as opposed to {\it nuclear} masses for the calculation of the Q-value in the above expression. The interacting nuclei in a stellar plasma are fully ionized. For example, the temperature range from 15 MK (Sun) to 300 MK (classical novae) corresponds to an average thermal energy of 1.3 $-$ 26 keV. The ionization energy of a 1s electron, for example, in carbon, neon, or argon is 0.3 keV, 0.9 keV, or 3.2 keV, respectively \cite{HUANG1976243}. Therefore, most nuclei taking part in the thermonuclear burning possess few, if any, bound electrons. Assuming that the interacting nuclei in a stellar plasma are fully ionized, the procedure of adopting {\it atomic} masses as opposed to {\it nuclear} masses for the calculation of Q-values disregards the difference in total electron binding energies. It will be shown that this effect can change thermonuclear reaction rates by significant amounts. As an example, the rate of the $^{36}$Ar(p,$\gamma$)$^{37}$K reaction will be discussed at temperatures between $30$~MK and $1$~GK.

This report is not concerned with the impact of atomic binding and excitation on the energy release in the laboratory study of nuclear reactions (see, e.g., Ref.~\cite{Christy:1961wr}). It is neither concerned with the shift in resonance energies caused by electron screening in the stellar plasma, which has to be taken separately into account (see, e.g., Refs.~\cite{Mitler:1977uq,iliadis15}).

Section~\ref{sec:qvalues} presents preliminaries. Results are discussed in Section~\ref{sec:example}. A summary is given in Section~\ref{sec:summary}.

\section{Q-values} \label{sec:qvalues}
Mass evaluations (e.g., Wang et al. \cite{Wang:2017cx}) present Q-values, or separation energies, based on {\it atomic} mass differences. Such Q-values are given by
\begin{equation}
Q_{at} = \Sigma m^i_{at} - \Sigma m^f_{at} 
\end{equation}
with $\Sigma m_{at}^i$ and $\Sigma m_{at}^f$ denoting the sum of atomic masses before and after the interaction, respectively. The quantity needed for computing center-of-mass resonance energies from measured excitation energies is the Q-value based on {\it nuclear} masses
\begin{equation}
Q_{nu} = \Sigma m^i_{nu} - \Sigma m^f_{nu} 
\end{equation}
Atomic and nuclear masses are related by
\begin{equation}
m_{at}(A,Z) = m_{nu}(A,Z) + Z m_e - B_e(Z)
\end{equation}
where $A$ and $Z$ denote the mass number and atomic number, respectively, $m_e$ is the electron rest mass, and $B_e(Z)$ is the total electron binding energy in the neutral atom of atomic number $Z$. A positive sign is assigned to the binding energy, $B_e$. Consequently, the Q-values based on nuclear and atomic masses are related by
\begin{equation}
\label{eq:qan}
Q_{nu} = Q_{at} + (\Sigma B_e^i - \Sigma B_e^f)
\end{equation}
where $\Sigma B_e^i$ and $\Sigma B_e^f$ are the sum of total electron binding energies before and after the interaction, respectively.

\begin{figure}[hbt!]
\includegraphics[width=0.9\columnwidth]{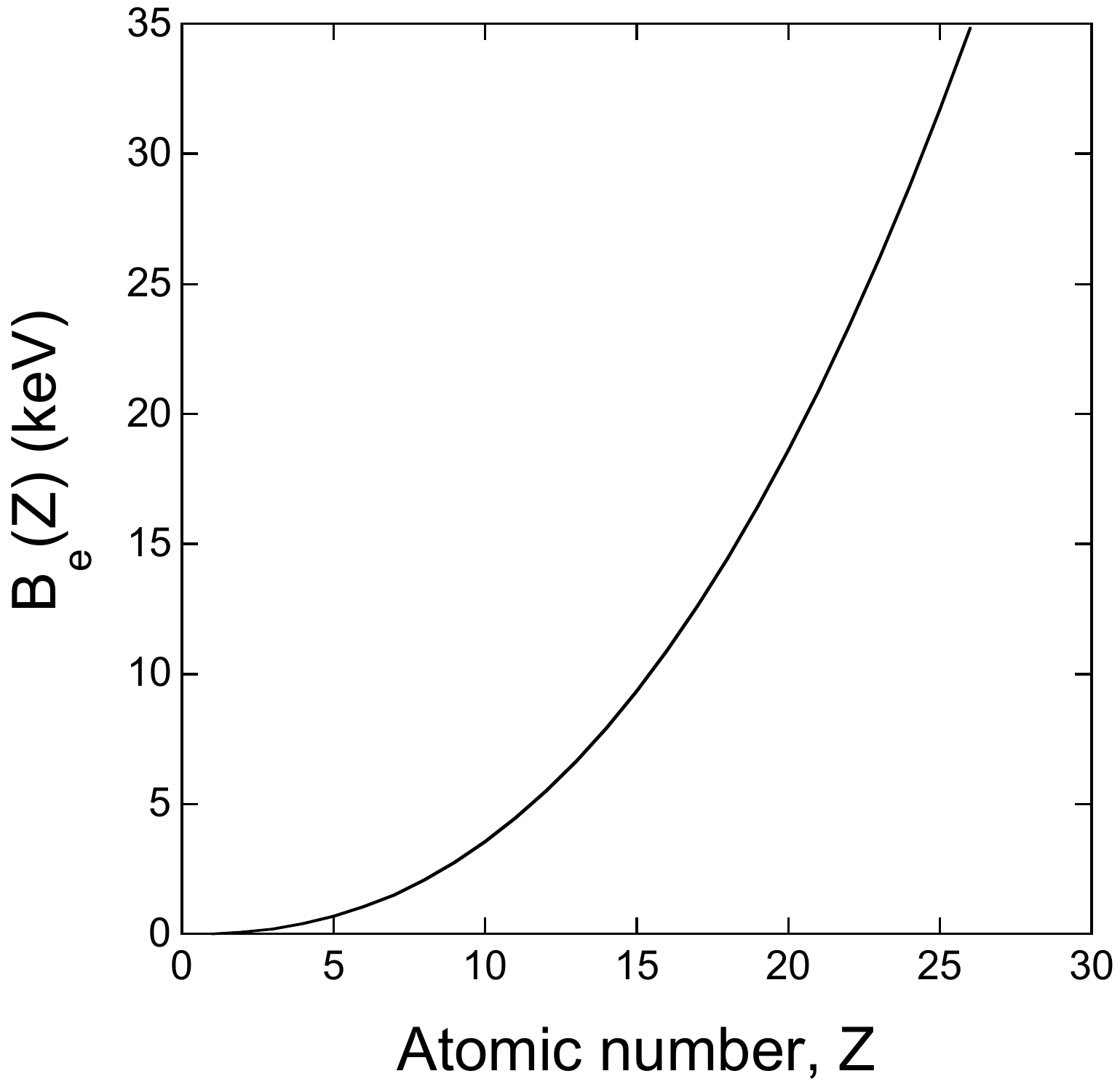}
\caption{\label{fig:binding} Total electron binding energy versus atomic number, $Z$. The graph is computed using the approximate expression from Lunney, Pearson and Thibault \cite{RevModPhys.75.1021}, given by Equation~(\ref{eq:lunney}), which is based on the calculations of Huang et al. \cite{HUANG1976243}.
}
\end{figure}

The total electron binding energy for an atom of given atomic number, $Z$, can be approximated by \cite{RevModPhys.75.1021}
\begin{equation}
\label{eq:lunney}
B_e(Z) = 14.4381 Z^{2.39} + 1.55468 \times 10^{-6} Z^{5.35} ~ \textrm{eV}
\end{equation}
which is based on the neutral-atom electron binding energies calculated by Huang et al. \cite{HUANG1976243} using the
relaxed-orbital relativistic Hartree-Fock-Slater formalism. This expression is plotted in Figure~\ref{fig:binding}. It can be seen that the binding energy in a neutral atom increases steeply for increasing atomic number: it amounts to $3.5$~keV for neon, $7.9$~keV for silicon, and $14.5$~keV for argon. In other words, disregarding the difference in total electron binding energies will introduce a significant bias in the calculation of the center-of-mass resonance energy from the excitation energy and the Q-value.

Total electron binding energies cannot be easily measured directly and, therefore, it is difficult to estimate the accuracy of Equation~(\ref{eq:lunney}). Table IV of Ref.~\cite{HUANG1976243} compares the calculated electron binding energies for given orbits to experimental values. The good agreement shows that Equation~(\ref{eq:lunney}) most likely estimates total electron binding energies with an uncertainty much smaller than the derived electron binding energy differences. Consequently, this correction should be taken into account, as illustrated in the following section.

\section{Example}\label{sec:example}
As an example, the thermonuclear rate of the $^{36}$Ar(p,$\gamma$)$^{37}$K reaction will be considered. Between temperatures of $70$~MK and $800$~MK, the rate is dominated by a single resonance near a laboratory energy of $\approx$320~keV (see Fig.~5 of Ref.~\cite{Iliadis:1992cd}). At lower temperatures, the direct capture process dominates the total rate. 

The resonance was first observed by Iliadis et al. \cite{Iliadis:1992cd}. The most recent value for the resonance strength is $\omega\gamma$ $=$ $(7.0\pm1.0)\times10^{-4}$~eV \cite{Mohr:1999tv}. The excitation energy of the corresponding $^{37}$K compound level is $E_x$ $=$ $2170.18\pm0.13$~keV \cite{DeEsch:1988wk}. The tabulated Q-value of the $^{36}$Ar(p,$\gamma$)$^{37}$K reaction, which is based on the difference in {\it atomic} masses, amounts to $Q_{at}$ $=$ $1857.63\pm$0.09~keV \cite{Wang:2017cx}. Based on this input, the center-of-mass resonance energy that would be adopted by most practitioners is $E_{r,at}^{cm}$ $=$ $E_x - Q_{at}$ $=$ $312.55\pm0.16$~keV. However, according to Equation~\ref{eq:qan}, the correct center-of-mass resonance energy based on the {\it nuclear} mass difference is
\begin{equation}
\begin{aligned}
E_{r,nu}^{cm} & = E_x - (Q_{at} + \Delta B_e) = E_{r,at}^{cm} - \Delta B_e \\ 
& = 314.53\pm0.16~\mathrm{keV}
\end{aligned}
\end{equation}
where $\Delta B_e$ $\equiv$ $\Sigma B_e^i$ $-$ $\Sigma B_e^f$ $=$ $-1.98$~keV is the total electron binding energy difference before and after the interaction. Therefore, disregarding the electron binding energy incorrectly reduces the resonance energy by about $2$~keV, which significantly exceeds the energy uncertainty.

Thermonuclear reaction rates can be very sensitive to resonance energy shifts of this magnitude, because the energy enters exponentially in the rate expression. The reaction rate per particle pair, in units of cm$^3$mol$^{-1}$s$^{-1}$, for a single isolated resonance is given by \cite{iliadis15}
\begin{equation}
\left<\sigma v\right> = \left( \frac{2\pi}{m_{01}kT} \right)^{3/2} \hbar^2 e^{-E_r^{cm}/(kT)} \omega\gamma
\end{equation}
where $m_{01}$ is the reduced mass of the $^{36}$Ar $+$ $p$ system, $k$ is the Boltzmann constant, and $T$ is the temperature. The ratio of the rate calculated with the resonance energy based on the nuclear mass difference and the rate based on the atomic mass difference is
\begin{equation}
\label{eq:defR}
R \equiv \frac{\left<\sigma v\right>_{nu}}{\left<\sigma v\right>_{at}} = \frac{e^{-(E_{r,at}^{cm}-\Delta B_e)/(kT)}}{e^{-E_{r,at}^{cm}/(kT)}} = e^{\Delta B_e/(kT)}
\end{equation}
where the small changes in the reduced mass and the de Broglie wavelength (which enters in the derivation of the measured resonance strength) can be safely disregarded. 

\begin{figure}[hbt!]
\includegraphics[width=0.9\columnwidth]{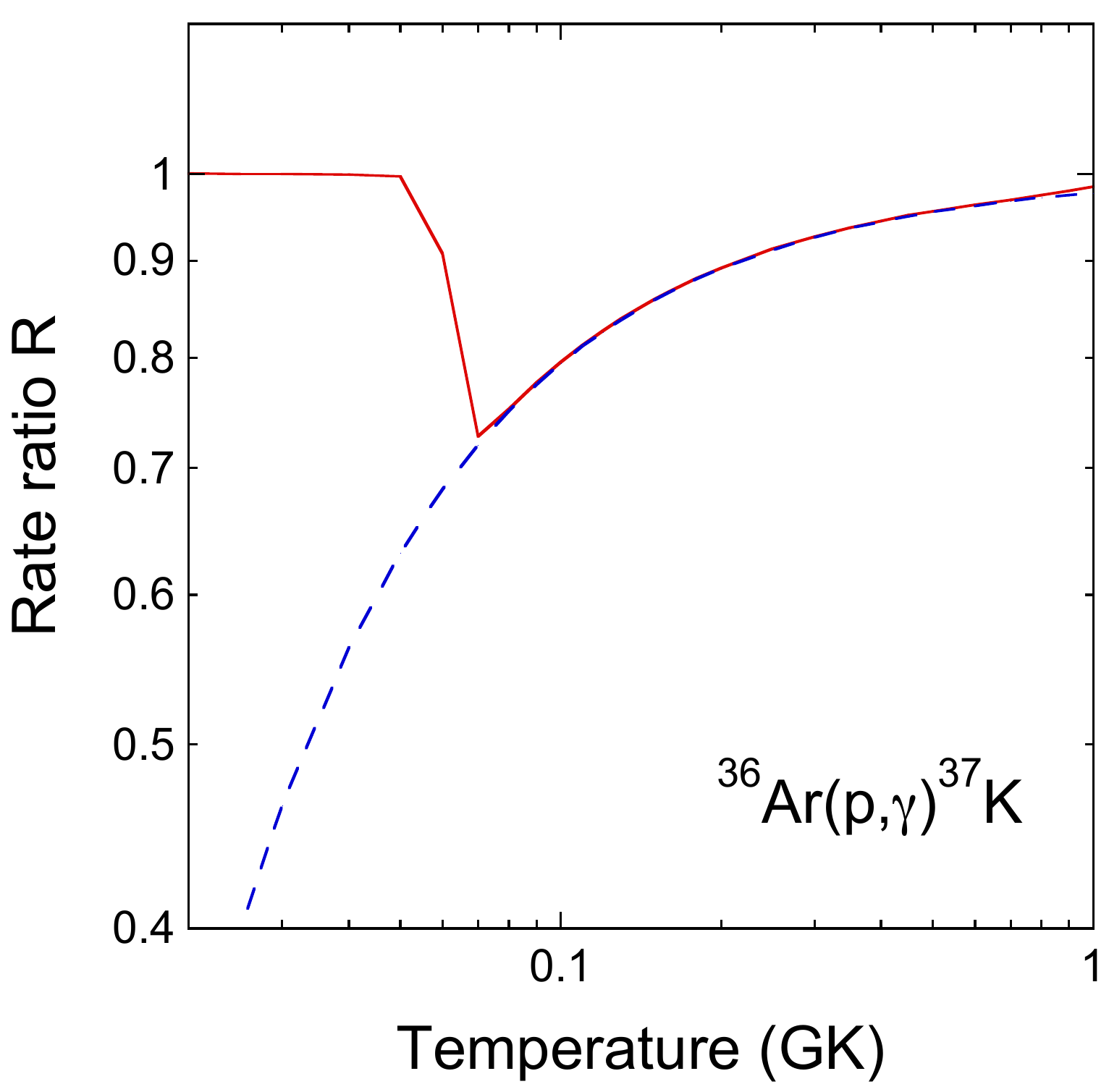}
\caption{\label{fig:result} (Color online) Ratio of $^{36}$Ar(p,$\gamma$)$^{37}$K reaction rates obtained with nuclear masses and atomic masses for calculating the center-of-mass resonance energy; the quantity $R$ is defined in Equation~(\ref{eq:defR}).  (Solid line) Rate ratio that takes all resonant and nonresonant contributions into account \cite{Iliadis:2010kva}. (Dashed line) Rate ratio for the $E_r^{c.m.}$ $=$ $315$~keV resonance alone, according to the expression $R$ $=$ $\exp{(-0.0230}/T_9)$ (see text).
}
\end{figure}

For the $315$~keV resonance in $^{36}$Ar(p,$\gamma$)$^{37}$K, one finds $R$ $=$ $\exp{(-0.0230}/T_9)$, where $T_9$ is the temperature in units of gigakelvin. This expression is plotted in Figure~\ref{fig:result} as the dashed line. It can be seen that the systematic bias introduced when atomic instead of nuclear masses are used can be significant. For example, at $70$~MK\footnote{The plasma ionization depends on the temperature, density (because of pressure ionization), and composition of the gas. A software instrument that computes the ionization fractions using the Saha equation is provided in Ref.~\cite{timmes}. For example, at $T$ $=$ $70$~MK one finds that $^{36}$Ar and $^{37}$K atoms are fully ionized for all densities and compositions of interest to thermonuclear burning.} the difference amounts to $\approx$40\%, and becomes larger with decreasing temperature. 

Besides the $315$~keV resonance, the direct capture process also contributes to the total $^{36}$Ar(p,$\gamma$)$^{37}$K rates below $1$~GK. The ratio of the total rates obtained with nuclear masses and atomic masses for calculating the center-of-mass resonance energy is shown as the solid line in Figure~\ref{fig:result}. As expected, it closely follows the dashed line above a temperature of $70$~MK, where the $315$~keV resonance dominates the total rate. For lower temperatures, the rate ratio is close to unity because the direct capture process dominates the total rate and the variation in the resonance energy is inconsequential.

\section{Summary}\label{sec:summary}
This work discussed the systematic bias introduced by adopting tabulated Q-values based on atomic mass difference for the calculation of center-of-mass resonance energies. This procedure, which is widespread in the literature, provides erroneous results since it does not account for the difference in total electron binding energies before and after the interaction. As an example, the thermonuclear rates of the $^{36}$Ar(p,$\gamma$)$^{37}$K reaction were discussed. In this case, the total electron binding energy difference causes a $2$-keV shift in the energy of the lowest-lying resonance, resulting in an erroneous increase of the reaction rate by $\approx$40\% near $70$~MK. The effect described in the present work is negligible for reactions involving light nuclides (p, d, t, $^{3}$He, $^{4}$He). However, it becomes noticeable for nuclides heavier than oxygen and increases in magnitude with increasing atomic number. Additionally, the effect will be more pronounced in $\alpha$-capture reactions compared to proton-induced reactions.

\begin{acknowledgments}
I would like to thank David Little for comments on the manuscript and Frank Timmes for his help with calculating ionization fractions in a plasma. This work was supported in part by the U.S. DOE under contracts DE-FG02-97ER41041 (UNC) and DE-FG02-97ER41033 (TUNL).  
\end{acknowledgments}



\bibliography{paper.bib}
\end{document}